# Gender Disparities in Nobel Prize Winning Labs. A Look into Glass Ceilings, Maternal Walls and Bottlenecks.

## Moaraj Hasan[1]

**Affiliation: 1 -** ETH Zurich
Corresponding Email: Mohasan@ethz.ch

## Abstract

In this study, headcounts of all personnel in Nobel Prize winning labs were collected and sorted by gender. These results are used to determine gender representation of graduate students in elite institutions on the pipeline towards higher academic positions. Larger gender disparities are seen in physics and physical chemistry labs and are reduced in biologically focused labs. These differences are greater in Nobel Prize winning institutions when compared to the USA and EU averages. The gender bias in hiring during the transition between post doctoral fellow and junior faculty seems to be the bottleneck for women; exacerbated by family formation. Women who surpass this hurdle achieve tenure at the same rate and do not perform any worse than men in such fields. Reduced participation in mathematically intensive fields can also be traced to propensities of girls' preferences to deviate from them as early as kindergarten. As such, gender disparity may not be due to recalcitrant pernicious attitudes of individuals towards women. Accumulated advantages such as societal expectation, asymmetric teaching efforts and acceptance of men to enter certain fields, over decades, may put men in more favorable positions while competing for placement in prestigious labs. However, these factors are sensitive to cultural shifts and show generational effects indicating the possibility of equality facilitated by sociocultural shifts in expectation for young women.

# I. The Big Picture

Despite much progress towards equal representation women still lag in executive positions in business and full professorship in elite academic institutions. Waterloo, the city of my Alma mater, leads Canada in gender inequality in terms of economic security, leadership, health, personal security, and education [8]. In Europe, female university graduates outnumber their male colleagues, but represent only 10% of the rectors of universities [12]. Aside from incidences like Tim Hunt recently vacating his honorary post at University College London for making untoward statements about female scientists, attitudes towards women in science are slowly changing [5, 18]. The growth of women entering academia outpace men by 2.3% except in the highest echelons of tenured professors, deans and presidents [12]. Doctorates awarded to women are close to parity in fields such as microbiology with large leakage of women in subsequent postdoctoral fellowships and assistant professors on the pipeline to full professorships [7].

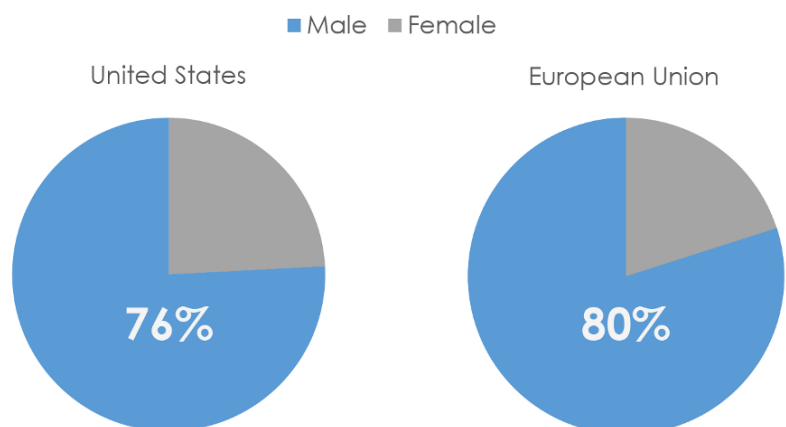

Figure 1: Figure 1: Gender Distributions of men and women fulfilling full time professor positions the United States [15] and the European Union [12] show women make up only a fifth of all positions

In this study, personnel data collected from Nobel Prize winning labs in the fields of medicine, chemistry and physics show women fulfill technician positions on par with men but take fewer PhD and postdoctoral positions. Although technicians are essential to functioning of labs, their work is infrequently given co-authorship or even acknowledged in publications. Publications are the fundamental metric for gauging academic success and hinder technicians from moving into professor positions without further graduate work [19]. Depending on the field, as little as 5.1% of all PhD recipients end up with tenure, and non-tenure track faculty positions[20]. Both men and women struggle to enter junior professor positions but the aggregate data hides the increased difficulty women experience over men [20].

The data collected for this article was taken from the laboratory webpages or publicly accessible directories of students and supervisors. A large caveat of using this set is that many lab websites are not updated regularly. To ensure only recent data is used, lab alumni before 2008 were not tallied. A 8 year mark is the upper limit of most doctoral appointments. Gender was determined using the following techniques.

1. Picture Provided on Website 2. Mention of Mr/Ms in directory or biography on website 3. LinkedIn or Facebook 4. Statistical determination of most probably gender using https://gender-api.com/

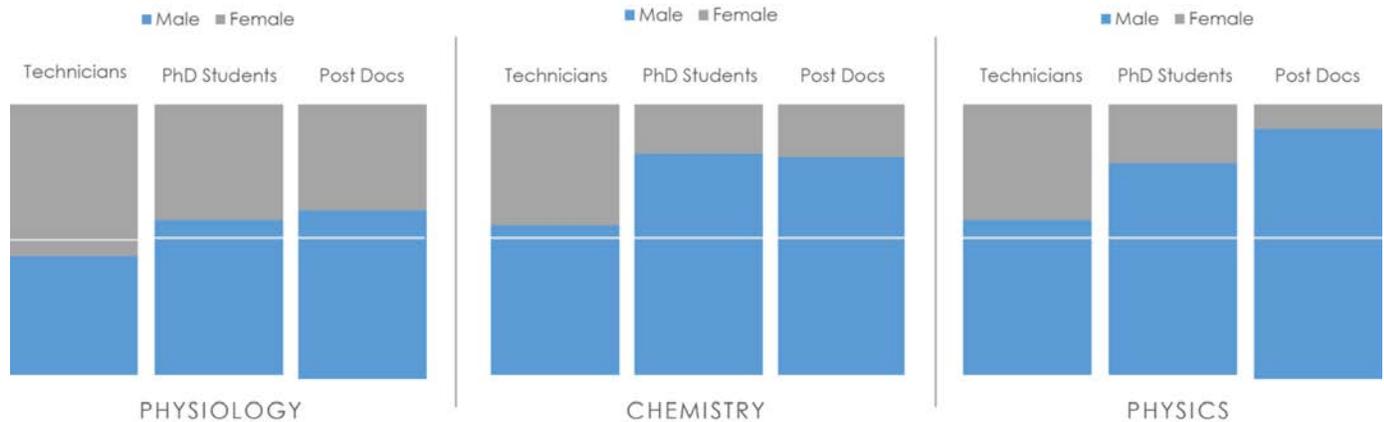

FIGURE 2: GENDER DISTRIBUTION OF MEN AND WOMEN IN TECHNICIAN, PHD AND POSTDOCTORAL FELLOWSHIPS IN LABS THAT HAVE BEEN AWARDED THE NOBEL PRIZE IN PHYSIOLOGY, CHEMISTRY AND PHYSICS RESPECTIVELY. WOMEN AND MEN SHOW SIMILAR PARTICIPATION IN TECHNICIAN ROLE

Figure 2 illustrates the overall gender distribution of all staff in institutes and laboratories that are currently run by a Nobel Prize winning principal investigator. Women are underrepresented in doctoral and post-doctoral positions in a large majority of Nobel Prize winning labs reducing their overall opportunity to contribute towards research and innovation; and given the different perspectives that women bring, the quality of research and innovation suffers as well [12]. If detriments in research quality from excluding women are known then why is it that this effect is so pervasive in the meritocratic realms of academia?

## II. Glass Ceilings and Maternal Walls

Nationwide tallies of female full professors in North America and Europe clearly indicate gender discrimination against women for faculty positions with increasing rates at more prestigious institutions [12, 15]. This discrimination may concomitantly be present at earlier stages of their scientific careers. A randomized application trial of otherwise identical CVs show substitution of male name with female name receive fewer call backs for hiring decisions [11]. A new study showing 71% of Postdoctoral fellows are picked from 16 elite institutions worldwide [1]. If the method for upward mobility in academia is to be a graduate student or postdoctoral fellow in such a school, the disparity of women at this level may explain the concentration of men in higher academic positions.

A PhD is an entry point into further academics or industry whereas post-doctoral fellowships emphasize grant writing and lab management skills that are critical for establishing a junior faculty's research program [13]. Scarcity of tenure tracked positions place a difficult choice for students to compete in saturated fields or become highly specialized further removed from the industrial canon [13]. Whereas hiring decisions for assistant professorships seem to be subject to gender bias, subsequent tenure appointments are given equally to men and women indicating the initial hiring from the postdoc pool is a bottleneck in the pipeline to tenure track faculty positions[6]. Figure 2 indicates postdoctoral hiring has a comparable gender bias as the hiring of entry level faculty positions seen in Figure 1. Although faculty hiring is much more extensive review of candidates, the combination of discrimination at postdoctoral and junior faculty hiring levels pose daunting challenges to women in an age where family establishment may also be an expectation[6].

Starting a family at this critical period can be disastrous to a woman's ability to establish a work-family balance before entering academics. The married mothers of children who are too young for school are 35% less likely to get tenure-track jobs compared with married fathers of young children. However, unmarried childless women are 4% more likely to get tenure-track jobs than are unmarried childless men. At this professional turning point, family formation probably explains why many female scientists don't get tenure-track jobs. Only 3% of female graduate students had access to at least six weeks of unrestricted leave, 58% for female faculty suggest family formation and children compound the effects of the hiring bottleneck en route to faculty jobs [10].

## III. Nobel Prize in Physiology and Medicine

Physiology and Medicine labs are the most egalitarian in terms of gender distributions of PhD/postdocs. Women Lag by 14%/23% on average in the labs surveyed as compared to their counterparts in chemistry and physic which have a 64%/63% and 52%/82% deficit. Biomedical and psychological research labs show equal entry into graduate school across North America and 14% calculated disparity may be a results of the small sample or a conditions in elite labs [9]. Greater prevalence in women in the field encourages prospecting graduate students to continue through modelling effects [16]. Regulations limiting the maximum hours per week for medical residents and increasing maternal recovery allowances have permitted many more women to enter the field of medicine [14]. MD postdoctoral fellows and MD/PhD students make up a large cohort of female entrants in North American institutes because individuals exist at a tipping point where gender bias no longer outweighs the strength of credentials [3

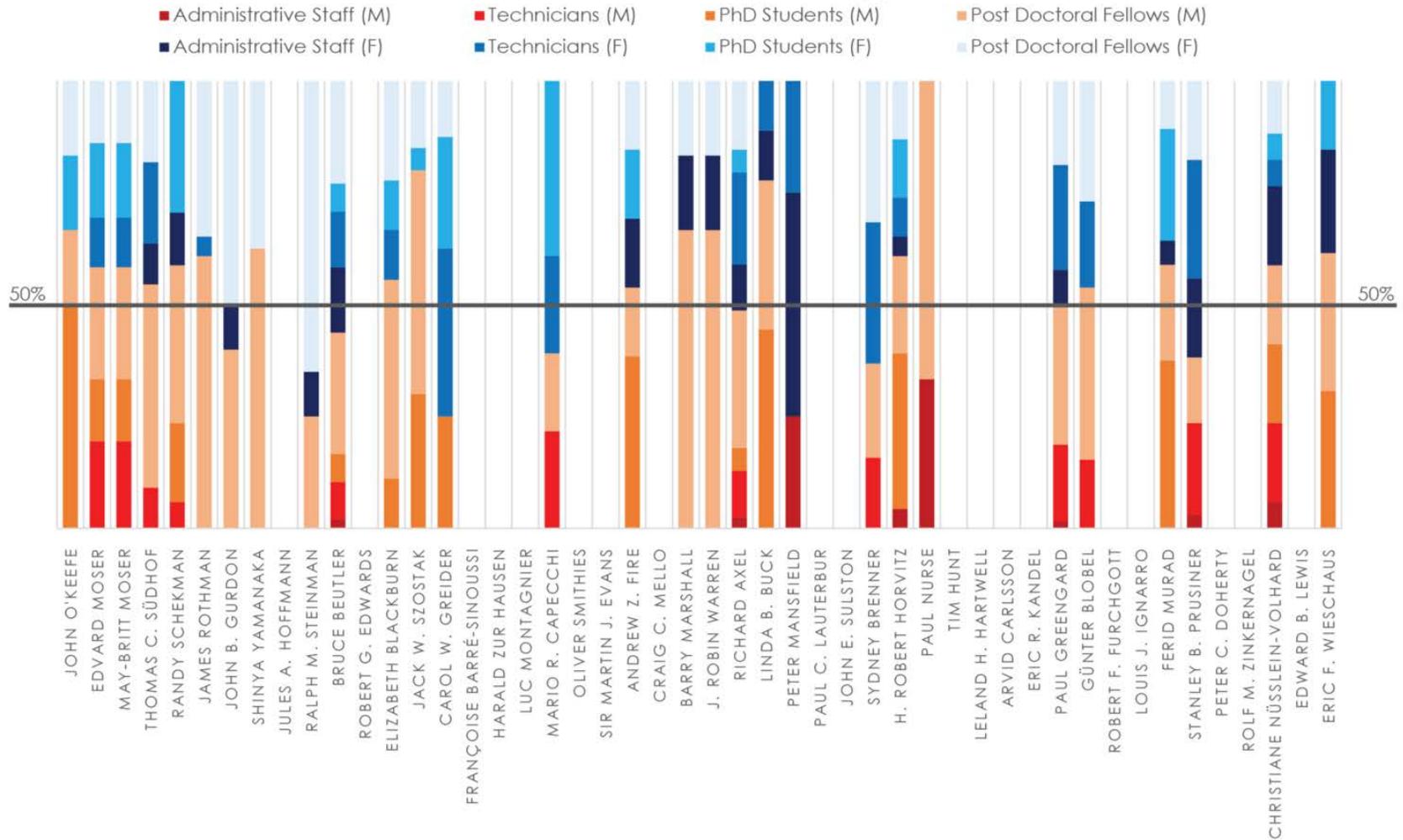

FIGURE 3: DISTRIBUTION OF MALES AND FEMALES IN TECHNICIAN, PHD AND POSTDOC POSITIONS IN LABS THAT HAVE WON THE NOBEL PRIZE IN PHYSIOLOGY. LABS THAT DID NOT HAVE PUBLICLY AVAILABLE DIRECTORIES AT THE TIME OF PUBLICATION OR HAD A PRINCIPAL

## IV. Nobel Prize in Chemistry

The Nobel Prize in chemistry exists on a spectrum from biochemical signalling, biomolecule characterizations to physical chemistry research such as rates of electron transfer reactions and applied physics work that produced super resolution microscopy. Roughly speaking, heavily biochemical labs show F/M ratios higher than physical chemistry groups which are still predominantly male with organic chemistry, analytical chemistry falling in the middle of the spectrum. However the small sample size and subjective qualification prevent specific rankings.

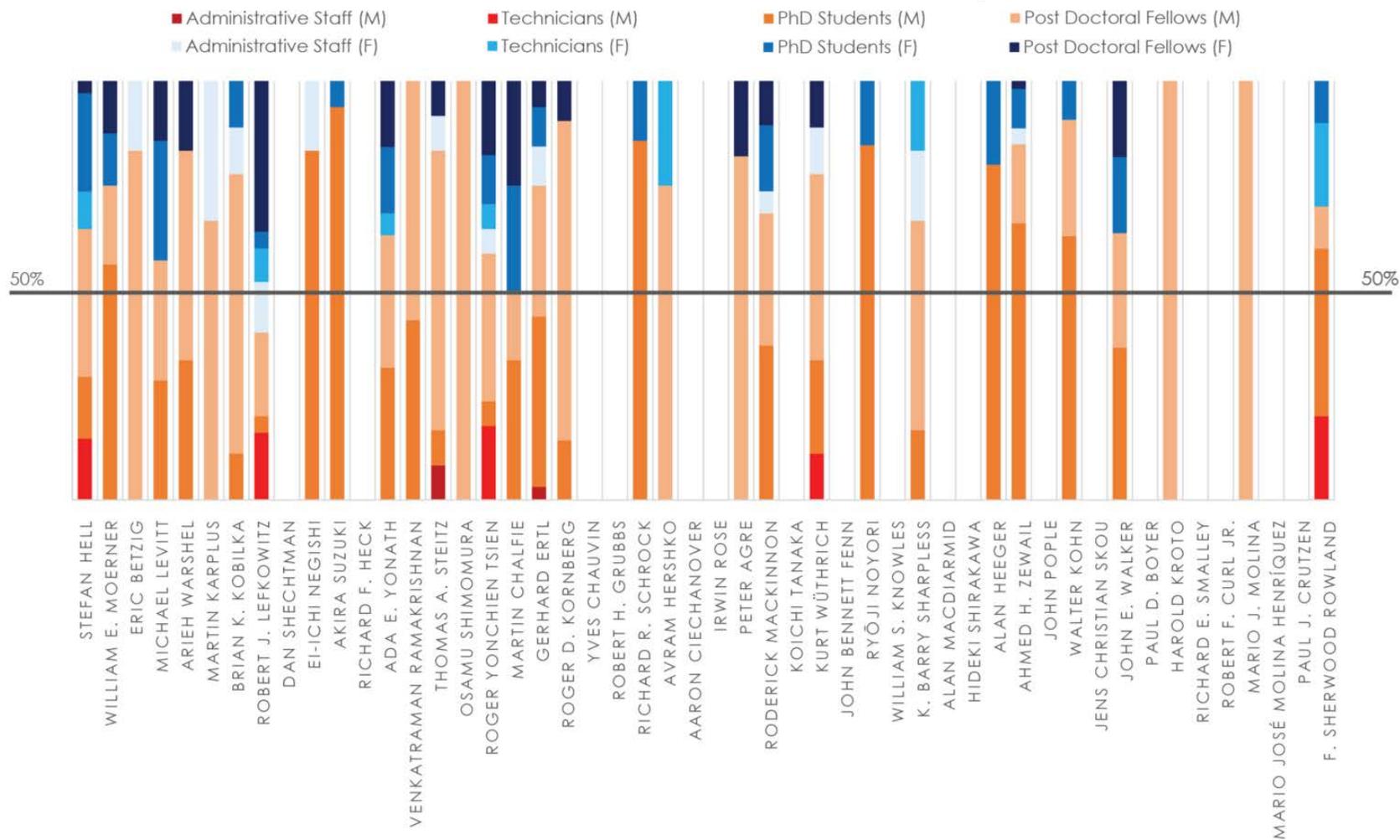

FIGURE 4: FIGURE 4: DISTRIBUTION OF MALES AND FEMALES IN TECHNICIAN, PHD AND POSTDOC POSITIONS IN LABS THAT HAVE WON THE NOBEL PRIZE IN CHEMISTRY. LABS THAT DID NOT HAVE PUBLICLY AVAILABLE DIRECTORIES AT THE TIME OF PUBLICATION OR HAD A PRINCIPAL INVESTIGATOR NO

# V. Nobel Prize in Physics

The data set for Nobel Prize winners in physics is sparse compared to chemistry and physiology because the lag between discovery is the largest and growing in the field of physics[17]. Professors that have taken an honorary position or have recently passed away are exempt from this analysis in order to portray demographics no older than 7 years(upper limit on most graduate appointments) [17]. Moreover, physics tends to have the smallest labs in which theoretical work and can be accomplished with fewer individuals whereas biological research is much more labor intensive [7]. The smaller number of open positions with significantly larger male participation in the field [9] produce the astonishing gender disparities from high school to postdoctoral fellowships in Nobel Prize winning labs[3].

*L*eslie et al. demonstrate how the way ability is viewed in different fields can correlate with the degree to which women are represented. Men dominate fields where ability is considered to be innate, such as philosophy and physics. Whereas, women are well-represented in fields that are labour-intensive, such as molecular biology and psychology, where effort and persistence are greatly valued [7].

Gender differences in attitudes toward and expectations about math careers and ability (controlling for actual ability) are evident by kindergarten and increase thereafter. This leads to lower female propensities to major in math-intensive subjects in college but higher female propensities to major in non-math-intensive sciences, with overall STEM majors at 50% female for more than a decade [2]. This finding accounts for gender sorting into Physics and other STEM fields at the high school and undergraduate level suggesting pre-college factors are primary determinants of girls' choice. Differences in expectations from peers and teachers can lead to subtle advantages for boys that choose their paths early and create an accumulated knowledge gap incrementally over decades [6]. After the completion of bachelor's degree women show leakages in continuation to phDs and gender bias at the postdoctoral level leads to a smaller pool of female researchers [3].

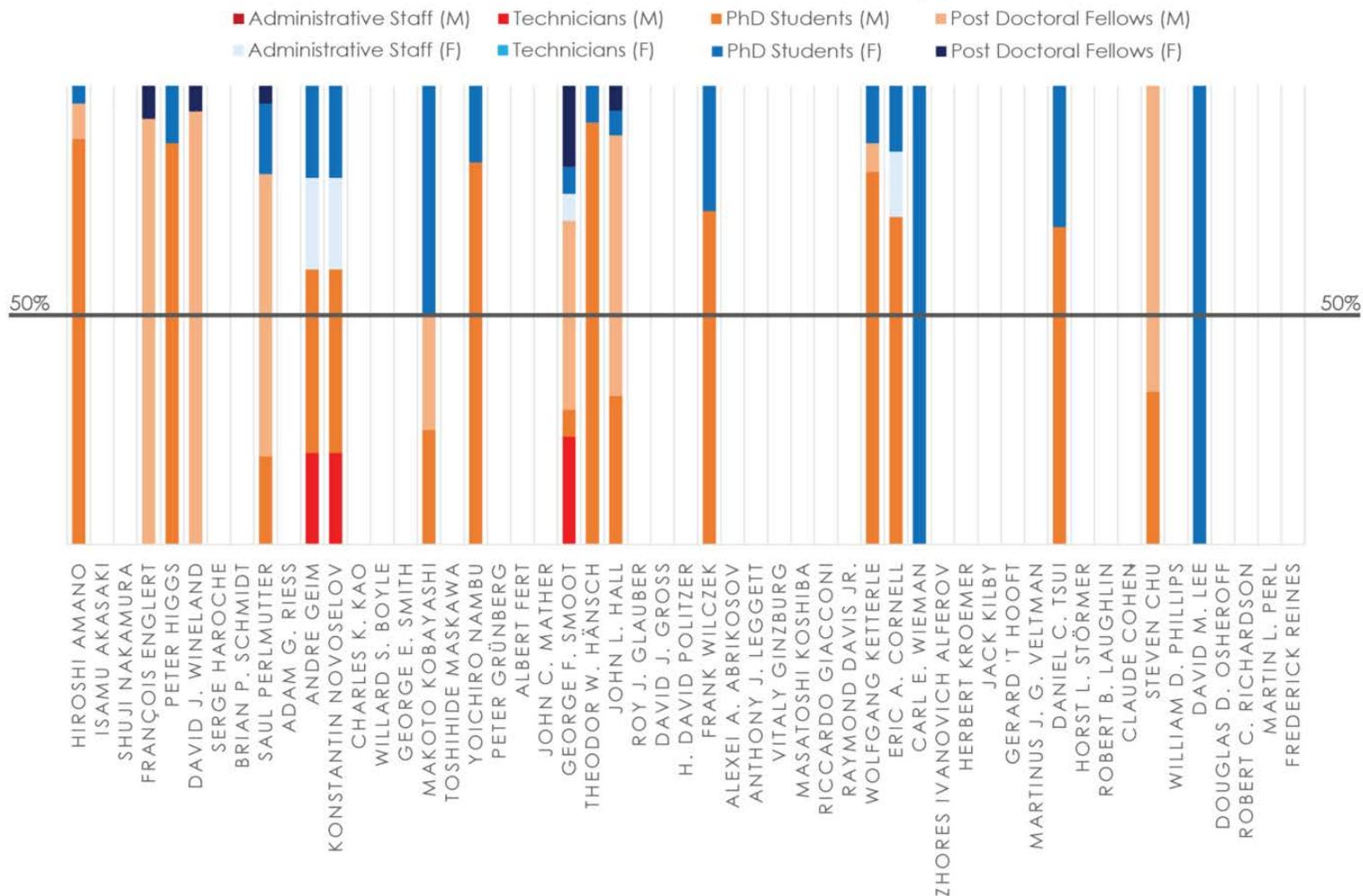

FIGURE 5: DISTRIBUTION OF MALES AND FEMALES IN TECHNICIAN, PHD AND POSTDOC POSITIONS IN LABS THAT HAVE WON THE NOBEL PRIZE IN CHEMISTRY. LABS THAT DID NOT HAVE PUBLICLY AVAILABLE DIRECTORIES AT THE TIME OF PUBLICATION OR HAD A PRINCIPAL INVESTIGATOR NO

# VI. The Next Line of Female Professors

The Nobel Prize captures the aspiration of many young scientists who seek out opportunities to work under tutelage of leading researchers. The numbers presented may be disheartening but fortunately, the factors that impede women from entering graduate and postdoctoral and ultimately professor positions in such elite institutes show large generational effects[12]. Older cohorts of scientists surveyed show much larger disparities in the participation of women in research [12]. At the moment, underrepresentation of women in STEM fields is strongly correlated to pre-college factors which alter the preference of women to major in certain fields and pursuing research paths rather than gender bias[2]. Research suggests differences in innate ability are unlikely to play a major role but one route to more equal representation across academic fields might be convincing both women and men that this is true [2]. Instead, early mentorship and encouragement from younger ages are primary determinants of whether a woman goes on in science and can facilitate large shifts towards equal representation of men and women at all levels of academia[4].

## Acknowledgements
Editing and Graphical Consultation:
Zuha Hasan, Novia Bui

# VII. Citations

1. **Systematic inequality and hierarchy in faculty hiring networks** A Clauset, S Arbesman, s, *Science* (2015), doi:10.1126/sciadv.1400005.http://advances.sciencemag.org/content/1/1/e1400005

2. **Women in Academic Science: A Changing Landscape.** Ceci SJ, Ginther DK, Kahn S, Williams WM (2014). Psychol Sci Public Interest 15: 75–141.

3. **Gender inequality in science** Andrew M. Penne Science 16 January 2015:**347** (6219), 234–235. [DOI:10.1126/science.aaa3781]

4. **Why are there still so few women in science? NY Times**http://www.nytimes.com/2013/10/06/magazine/why-are-there-still-so-few-women-in-science.html

5. **Sir Tim Hunt FRS and UCL** https://www.ucl.ac.uk/news/news-articles/0615/100615-tim-hunt

6. **Why So Slow?: The Advancement of Women**https://books.google.ca/books?hl=en&lr=&id=qikWHO0JecAC&oi=fnd&pg=PR9&ots=ByEjTYLerd&sig=Bd7F00wHXkVFQCr7b55ykJKdgCY&redir_esc=y#v=onepage&q&f=fals

7. **Expectations of brilliance underlie gender distributions across academic disciplines**http://www.sciencemag.org.proxy.lib.uwaterloo.ca/content/347/6219/262.abstract?ijkey=9e0fc358d1d08ddb385c5de173cdd423dc5c3d27&keytype2=tf_ipsecsha

8. **The Gender Gap in Canada's 25 Biggest Cities**https://www.policyalternatives.ca/publications/reports/best-and-worst-places-be-woman-canada-2015#sthash.FyN9BZyb.dpufhttps://www.policyalternatives.ca/publications/reports/best-and-worst-places-be-woman-canada-2015

9. **Philosophy in Disciplinary Perspective: Percentage of U.S. Ph.Ds awarded to Women in 2009**

10. **Do Babies Matter?: Gender and Family in the Ivory Tower**https://books.google.ca/books?id=-XMRAAAAQBAJ&lpg=PA59&dq=Do%20Babies%20Matter%3A%20Gender%20%26%20Family%20in%20the%20Ivory%20Tower&pg=PA58#v=onepage&q=Do%20Babies%20Matter:%20Gender%20&%20Family%20in%20the%20Ivory%20Tower&f=false

11. **The impact of gender on the review of the curricula vitae of job applicants and tenure candidates: A national empirical study.**Steinpreis, R. E., Anders, K. A. & Ritzke, D. *Sex roles* **41,** 509–528 (1999).

12. **She Figures 2012 Statistics and Indicators and Innovation Gender in Research**. EUROPEAN COMMISSIONhttp://ec.europa.eu/research/science-society/document_library/pdf_06/she-figures-2012_en.pdf

13. **The 2013 Canadian Postdoc Survey: Painting a Picture of Canadian Postdoctoral Scholars.** CAPS-ACSP and Mitacshttps://www.mitacs.ca/sites/default/files/caps-mitacs_postdoc_report-full_oct22013-final.pdf

14. **Trends in tenure for clinical MD faculty in US medical schools: a 25-year review.** Bunton, SA & Corrice, A. (2010). at <http://pdxscholar.library.pdx.edu/elp_fac/22/>

15. **Gender and Race of Tenured Factuly in the United States**http://iwl.rutgers.edu/documents/njwomencount/Faculty%20Diversity-3.pdf

16. **The Effect of Role Models on the Attitudes and Career Choices of Female Students Enrolled in High School Science**. Van Raden, Stephanie Justine, (2011). Dissertations and Theses. Paper 370.

17. **The Nobel Prize delay.** Physics Today. Francesco Becattini, Arnab Chatterjee, Santo Fortunato, Raj Kumar Pan, Pietro Della Briotta Parolo and Marija Mitrovic http://scitation.aip.org/content/aip/magazine/physicstoday/news/10.1063/PT.5.2012

18. **Intending to Stay: Images of Scientists, attitudes towards women and gender and influences on persistence among science and engineering majors.** Maryer Wyer. Journal of Women And Minorities in Science and Engineering

19. **Keeping the culture alive: the laboratory technician in mid-twentieth-century British medical research.** E.M Tansey. DOI: 10.1098/rsnr.2007.0035

20. **Paths to the Professoriate: Strategies for Enriching the Preparation of Future Faculty**. Donald H. Wulff, Ann E. Austin & Associates. http://depts.washington.edu/cirgeweb/wordpress/wp-content/uploads/2012/11/so-you-want-to-become-a-professor.pdf

# Data and Website Address

Graphs, Excel Files available upon request at Moaraj@moaraj.com